\documentclass[runningheads]{llncs}
\usepackage[T1]{fontenc}
\usepackage{longtable}
\usepackage{xcolor}
\usepackage{comment}
\usepackage{listings}

\usepackage{graphicx}
\begin{document}
\title{RAINO: Anchoring Agents in Reality \\
\large{A Systematic Review and Conceptual Framework for Realism in Agent-Based Modelling}}
\titlerunning{RAINO: Anchoring Agents in Reality}
%
 \author{ Lo\"is~Vanh\'ee$^{1,2}$ $^{[0000-0002-4147-4558]}$ 
 \and
         Melania Borit$^2$ $^{[0000-0002-1305-8581]}$
         \newline
 \email{lois.vanhee@umu.se}\\ 
 \email{melania.borit@uit.no}
       }

 \authorrunning{L. Vanh\'ee and M. Borit}

%
 \institute{Ume\aa{} Universitet; Ume\aa{}, Sweden \and UiT The Arctic University of Norway; Troms\o, Norway
}
%
\maketitle              
\begin{abstract}
Realism is a central yet seemingly under-theorized concept in Agent-Based Modelling. This paper presents a Systematic Literature Review, aiming to identify how realism is currently operationalized and demonstrated. The results show that realism is often poorly defined and lacks a consistent conceptual framework. A wide variety of methods are used to achieve and demonstrate realism, but explanations of whether and why these methods are appropriate for their intended purposes are generally limited. Building on this review, we introduce the Reality Anchor, Input, Output (RAINO) framework. RAINO identifies the key structures used to argue for realism in Agent-Based Models, consisting of Reality Anchors (e.g., empirical data, formal theory, expert knowledge, common-sense expectations) and their application as model Input or Output. RAINO broadens existing perspectives on how realism is framed. It explains why different assessors may evaluate the realism of a model in different ways, and it shows how this broader framing can lead to significantly different approaches to model development.

\end{abstract}

\section{Introduction}

The design of realistic models is a central concern in modelling in general and in Agent-Based Models (ABM) and Social Simulation in particular. Modelling often aims to reproduce real-world phenomena \cite{squazzoni2014social}, ranging from minimalist models that try to capture complex emergent behaviours \cite{epstein1996growing,schelling1971dynamic} to highly detailed models that attempt to reflect reality in depth \cite{barrett2008episimdemics}. Efforts to make ABMs and their components more realistic have attracted dedicated research communities \cite{jager2017enhancing} and frequently serve as motivation for proposing new models \cite{jensen2023dynamic,kammler2025modeling,sun2006prolegomena}. Even when ABMs are designed to explore fictional or hypothetical worlds, their designers often strive to keep the non-fictional elements as close to reality as possible \cite{bainbridge2018interstellar,gerdes2023commonsim}. In this sense, realism is typically the default assumption.

Despite its central role, the use of realism in ABMs (in the sense of creating "a realistic model"\footnote{This paper focuses on how to make ABMs closer to reality. Philosophical discussions surrounding realism, such as ``does reality exist?", were not addressed \cite{bhaskar2013philosophy,miller2015agent}.}) is most often colloquial and implicit. Realism is typically taken to mean that models "try to represent life as it really is" (as defined in the artistic sense by the Cambridge Dictionary \footnote{https://dictionary.cambridge.org/dictionary/english/realism}), without further specification. However, this intuitive understanding is problematic: from one ABM (or one paper) to another, there can be radical differences in how realism is intended, produced, demonstrated, and argued for \cite{ball2025assessing,chandan2020integration,kwon2013level,moss1998social,yao2024reinforcement}. In particular, multiple standard processes and methods applied during the modelling process can be argued for creating and/or demonstrating (some form of) realism, such as specification, representation, calibration, validation, and empirical grounding \cite{castella2005participatory,collins2024methods,mcculloch2022calibrating,sargent2010verification,scheller2019review,smajgl2014empirical,windrum2007empirical}. While there is an overlap, these concepts are different: stating "a validated model of..." is qualitatively different from stating "a realistic model of...". At this point, despite the centrality of the concept of realism to Agent-Based Modelling as a field, there is no comprehensive framework for capturing the concepts and implications tied to realism. Better seizing this concept is essential for enabling its effective and objective implementation and demonstration, which is important when modelling and exchanging information about our models, both within the ABM community and when engaging across fields and with stakeholders.

As a practical step towards developing a conceptualization of realism, this paper is dedicated to address the research question: "\textbf{What arguments have been used for showing that a given ABM is realistic?}". Methodologically, this paper addresses this question by applying a Systematic Literature Review (SLR) that compiles \textit{explicit} discourses about realism in ABM publications that explicitly mention realism in their titles (N=73 documents), organizing the data collection and analysis along 1) how realism is defined, 2) how realism is framed in the existing literature, including how realism is qualified by and qualifies other concepts; 3) what methods are used to make a model realistic and how these methods are used as arguments for showing the realism of the model; and 4) what methods are used for assessing the realism of a model based on its behavior and how these assessments are used as arguments for showing the realism of the model. 

As this SLR identifies the absence of thorough discussions, conceptualizations, and theories relative to realism, our paper then addresses this gap by compiling the argumentative structures deployed for showing the realism of models in an integrative framework that we call RAINO, for \textit{Reality Anchor, Input, Output}. 
RAINO introduces two key dimensions: \textit{Reality Anchors} i.e., the element that is used as an account of reality (e.g., data, theories, expert knowledge, processes, common-sense expectations), and \textit{Input/Output} i.e., whether these anchors relate to how the model is constructed (e.g., decision and interaction processes, context-specific elements, implementation methods) or how the model behaves (e.g., replication of certain properties and patterns of the anchor in the system's output). Moreover, integrating the plurality of arguments and assessments of similar models, RAINO integrates the \textit{subjective factors} commonly observed in arguments and assessments of the realism of models.

By integrating these essential yet often overlooked conceptual underpinnings, the RAINO framework allows for \textit{revisiting concepts, observations, paradoxes, and operational strategies that are core to Agent-Based Modelling research and practice}, including: 
why various actors tend to assess differently the realism of ABMs? How related concepts, such as validation, verification, calibration, and empirical grounding, relate to realism? How do KISS and KIDS modelling strategies relate to model realism \cite{edmonds2004kiss}? What are the ethical entanglements of realism in ABM design and use? Overall, we argue that RAINO provides the necessary foundations for supporting the development of further conceptualization and possible re-negotiation of now-uncharted yet central concept for Agent-Based Modelling that is realism, as well as enabling the development of effective practical approaches for producing realistic agent-based models and demonstrating their realism.



\section{Related work}

To our knowledge, no previous research has directly applied systematic methods for identifying and framing how realism is used in Agent-Based Modelling research, as performed by this paper. However, two previous lines of research are connected to the current activity.

Within the related field of Artificial Agents, Reynaud et al. (2018) \cite{frenchrealism} engage in a (non-systematic) review of aspects of realism and propose a framing of realism along three levels of proximity with reality: individual-level realism (defined as the replication of observed behaviors of specific individuals that are meant to be simulated), local-level realism (defined as the replication of expected real-world meso-level emerging properties, such as "every individual is expected to cook once per week", although there is no evidence or matching with evidence on who specifically cooks when), and aggregated-level realism (defined as the replication of general trends at the population level, without seeking to demonstrate proximity with specific individuals).

Giardini et al. \cite{giardini2023modeling} use a systematic method (Rapid Literature Review) on the more specialized case of realistic human behavior in the context of crises. Leaving aside a conceptualization of realism, this study identifies that, in the analyzed literature, realism is studied along psychological, behavioral, and social dimensions (e.g., a psychologically realistic model should integrate and/or demonstrate a proximity with psychological properties as observed in reality). This study also indicates that making models more realistic demands the collection of further data and the further involvement of relevant stakeholders, although it also suggests that further data collection may not lead to enhanced realism, because that data may become irrelevant in the concrete and specific context they operate in.

Besides research directly engaging with the concept of realism, a set of processes commonly carried as part of the modelling activity can be argued to relate to realism: calibration, validation, empirical grounding \cite{castella2005participatory,collins2024methods,mcculloch2022calibrating,sargent2010verification,scheller2019review,smajgl2014empirical,windrum2007empirical}. Despite their prevalence in ABM research, to our knowledge, no explicit discussion connecting these concepts to realism has been proposed thus far.

\section{Method}

\begin{figure}[t]
    \centering
    \colorbox{white}{
    \includegraphics[width=0.7\linewidth]{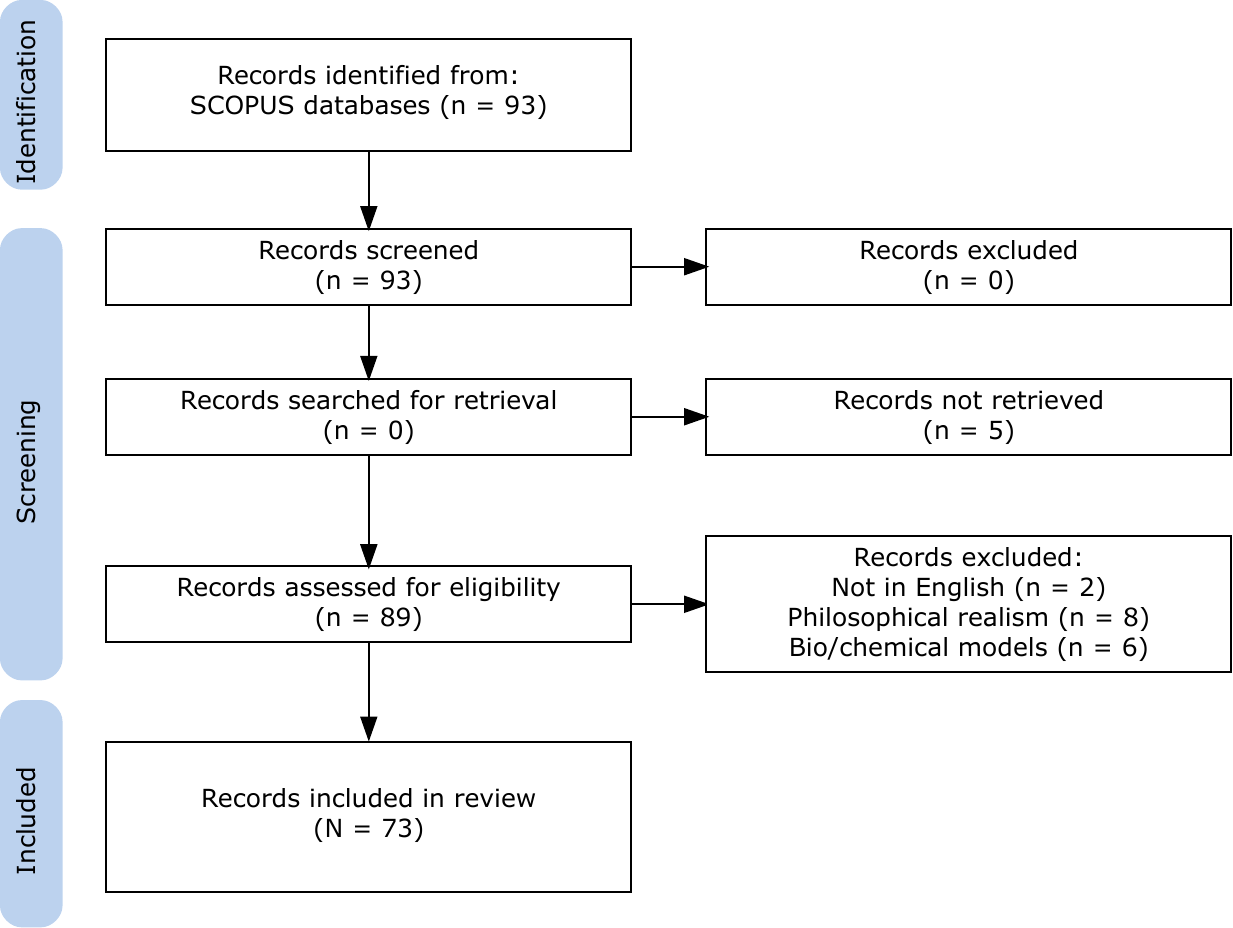}}
    \caption{PRISMA diagram of the selection process, generated using \cite{haddaway2022prisma2020}}
    \label{fig:prisma}
\end{figure}

To address our research questions, we applied a Systematic Literature Review (SLR) method.
The SLR process was conducted in accordance with established guidelines for systematic reviews in conservation and environmental science \cite{pullin2006guidelines}, as well as the PRISMA reporting standards \cite{moher2010preferred} and the recommendations in \cite{achter2024conduct}. 
These frameworks provided a structured approach to accurately sample and analyze the literature during the SLR. The PRISMA diagram summarizing the steps of this SLR is presented in Figure \ref{fig:prisma}.
The work unfolded in three stages. First, relevant publications were systematically and reproducibly gathered and selected; second, the chosen literature was examined qualitatively in-depth using content analysis and hierarchical coding; third, data visualizations and a summary of the results were created.

\vspace{2mm}
\noindent\textbf{Literature collection and selection}
We conducted our literature search using the academic search platform Scopus (www.scopus.com). In order to obtain a set of the most relevant documents, we applied the following process.
First, we identified a set of keywords that would allow us to find the literature most pertinent to the aim of the SLR. To identify documents referring to realism, we initially used the keyword "realis*". However, because this query retrieved false positives such as "realisation", we refined the keywords to "realism" OR "realist*". Second, we established a list of keywords related to ABMs, based on the keywords from \cite{vanhee2023thirty}. Due to the large number of retrieved papers, we restricted the query to agent-based/individual-based models/simulations and social simulation. We then searched for documents featuring the realism-related keywords in their title, abstract, author keywords, and/or publication venues. 
Due to the high number of retrieved documents (N=2141), we selected 20 articles at random and found that most of these articles only loosely related to realism. To retain the articles most likely to engage deeply with realism, we further restricted the query to documents containing markers of realism in their title. This refined search yielded n=93 documents, which was considered satisfactory for proceeding with the screening, with N=73 documents being retained for in depth analysis (Table 1 in Annexes). The search was performed on 01/04/2025. The final query was:

\noindent
\begin{lstlisting}
TITLE("realism*" OR "realist*") AND ((TITLE-ABS("agent-based model*" OR "agent-based simulation" OR "agentbased model*" OR "agentbased simulation" OR "individual-based model*" OR "social simulation")) OR (SRCTITLE("agent-based model*" OR "agent-based simulation" OR "agentbased model*" OR "agentbased simulation" OR "individual-based model*" OR "social simulation")) OR (AUTHKEY("agent-based model*" OR "agent-based simulation" OR "agentbased model*" OR "agentbased simulation" OR "individual-based model*" OR "social simulation")))
\end{lstlisting}

\noindent\textbf{Literature analysis}
A coding scheme was established in order to directly substantiate the research questions and was systematically applied to all of the documents kept for analysis, primarily by searching for all occurrences of the use of realism within the documents. As we focus on argumentative structures related to realism, our approach to coding sought to be as encompassing as possible. For example, a paper mentioning that method X may make a model more realistic was sufficient to mark that this paper related X as a method for seeking to create realism. Likewise, paper-specific descriptions were associated with overarching ABM concepts (e.g., "the value of the gamma factor was set to maximize the realism of the model" was coded as "calibration").

\section{Results}

\subsection{General patterns}

\begin{figure}[t]
    \centering
\colorbox{white}{
\includegraphics[width=0.85\linewidth]{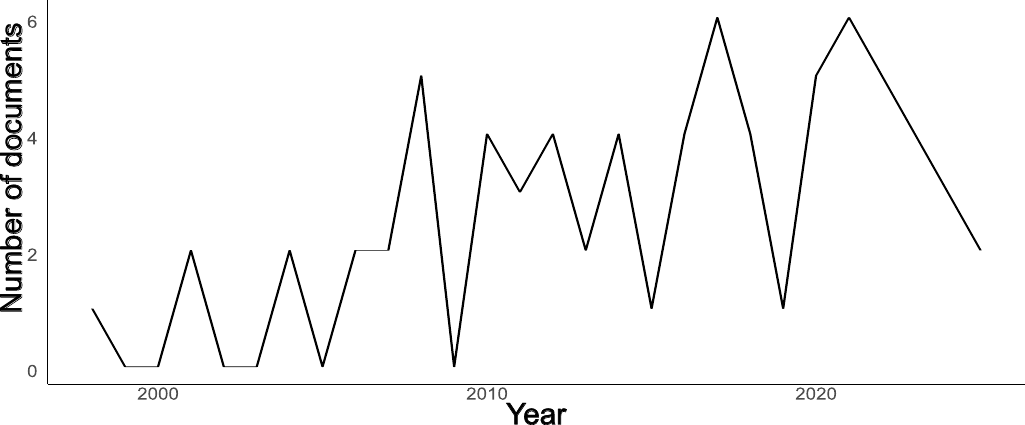}}
    \caption{Number of documents per year that were kept for in-depth analysis.}
    \label{fig:nb_docs_per_year}
\end{figure}

\begin{figure}[t]
    \centering

\colorbox{white}{
\includegraphics[width=0.9\linewidth]{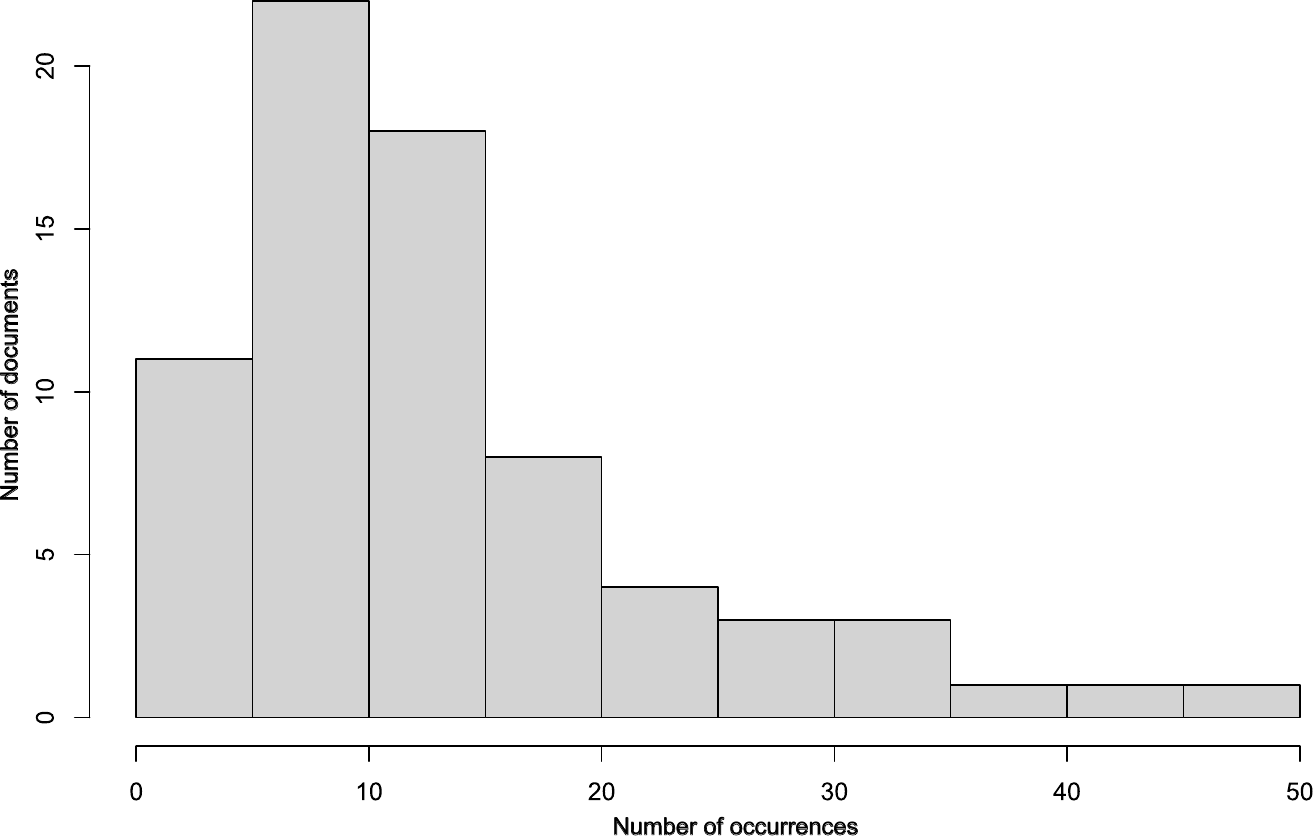}}
    \caption{Histogram of the documents featuring a number of occurrences of \textit{realism} and \textit{realist*} in their text.}
    \label{fig:histogram_frequency}
\end{figure}

\begin{figure}[t]
    \centering
    \includegraphics[width=0.65\linewidth]{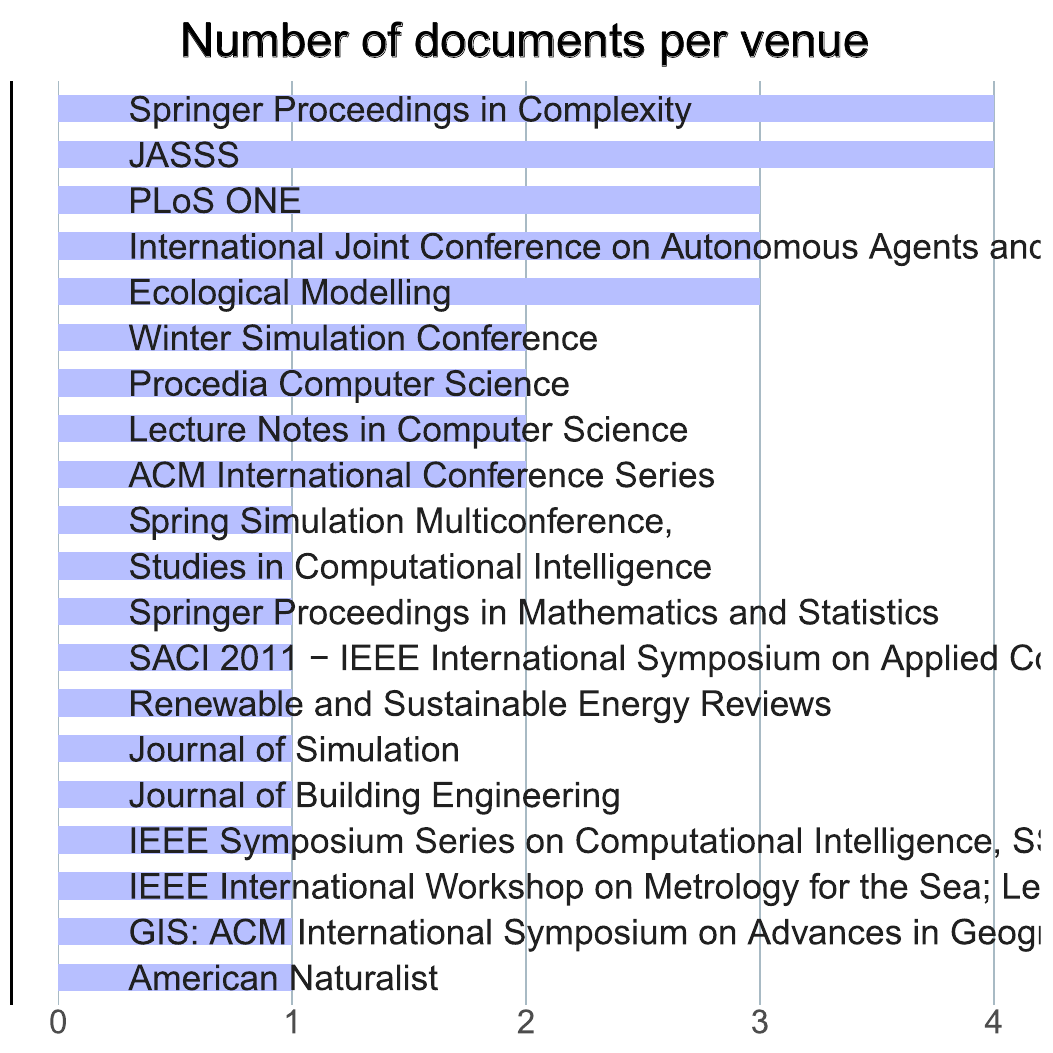}
    \caption{Top twenty venues having the largest number of documents that were kept for in-depth analysis.}
    \label{fig:nb_docs_per_venue}
\end{figure}

Figure \ref{fig:nb_docs_per_year} (see Annexes) summarizes the distribution of the documents retrieved for analysis by year of publication. The corpus spans from 1998 to 2025, and features a steady progression over time. Figure \ref{fig:nb_docs_per_venue} (see  Annexes) summarizes the top 20 publication venues based on the number of documents they have published, showing that this line of research is primarily published in general social simulation venues, followed by social simulation applications venues (e.g., ecological modelling, sustainable energy).
An overall analysis of the use of realism in the corpus highlights that \textit{the engagement of existing literature with realism is limited}. The SLR retrieved \textit{zero} papers dedicated to defining, framing, or surveying \textit{the use of realism \textbf{specifically in ABMs}}. When mentioned, realism was rarely introduced or discussed. While observed in the thorough reading of the documents, this observation was complemented by counting the number of uses of "realism" and "realist*" per text, listed in Figure \ref{fig:histogram_frequency} (see Annexes).
This shows a significant skew towards 10 uses of per document, with an average of 13.6 uses per document. As a measure of comparison, "simulation" and "model" were used 72 times more often.

\subsection{Defining and framing realism}

\noindent\textbf{Defining realism} Two documents provide explicit general definitions of realism in modelling/simulation: 
\textit{a realistic simulation must be one that closely approximates objective reality} \cite{ipto2008realistic}
and
\textit{the resemblance of the model to the real world} \cite{mercuur2018realistic}. One document provides a semi-general definition of realism:
\textit{a realistic simulation should look and behave like a real-world market} \cite{yao2024reinforcement}. Three definitions focus on specific perspectives of realism: behavioral realism defined as \textit{more complete representation of complex consumer behavior} \cite{ball2025assessing}; structural realism defined as \textit{a 
higher chance that our model captures the 
key aspects of an ecological system’s internal organization} \cite{grimm2016structural}; and organizational realism defined as having the \textit{attributes of the artificial organization of agents conform to empirical results for human organizational systems} \cite{hazy2004building}. The rest of the articles, while not defining realism, seemingly uniformly implicitly relied on the dictionary definition of realism (i.e., proximity with what is considered to be the reality).

Summing up, it seems that, despite being a specialized literature featuring realism in its title, the introduction of realism remains very rare (\(<10\%\) of the time) and limited (one sentence). Albeit subtle, the proposed definitions introduce meaningful differences in how realism can be considered. While all definitions align in framing realism as an indicator of proximity between a model and reality, divergences can be observed in the \textit{nature} of this proximity (e.g.,~resemblance, approximation, completeness in the sense of the number of aspects included in the model) and in how \textit{reality is attained} (as real world, objective reality, data, theories, expectations, resemblances).

\vspace{0.1cm}

\noindent\textbf{Framing realism}
Realism of ABMs is often framed in relation to other models. Numerous articles frame realism as a \textit{dimensional} construct. 
In this view, each model has a \textit{degree of realism} and can be more or less realistic than other models, including a theoretical maximum on ABM realism \cite{mercuur2018realistic}. Some articles expand this notion into realism as a \textit{multi-dimensional} construct. 
Usually, the model is said to be closer to reality along specific facets (e.g., the model can be realistic along its psychological dimension but produce unrealistic social dynamics), some of which were identified by \cite{giardini2023modeling}. One paper briefly suggests a radically different framing: ABMs can be realistic in regard to \textit{a given purpose} \cite{grimm2020odd}.

Realism of ABMs is often related to other concepts that can qualify or be qualified by realism. In our data, concepts that qualify realism include: psychological realism,
behavioral realism, 
social realism, 
qualitative realism,
graphical realism,
spatial realism, 
statistical realism, 
biological realism, 
ecological realism,
structural realism, 
model realism, 
and micro/meso/macro realism.
Realism qualifies concepts such as: models, simulations, mechanisms (e.g., diffusion mechanisms), data, population densities, statistical properties, variation, topologies, settings, benchmarks, case-specific properties (e.g., harvest regimes, fishing pressure; selection, evolutionary timescale, prices, market conditions, flow, pedestrian navigation paths), and assumptions. The extensiveness of these lists shows the versatility of the use of realism, which underlines a plausible high context-sensitivity of the concept. Although frequent and varied, the majority of these approaches to realism are expressed in less than a sentence, leaving ample space for interpretation.

\subsection{Making Agent-Based Models realistic}

The primary way ABMs and realism are related in the analyzed data is through describing methods for making the model more realistic (n=64, 88\%). A broad array of methods is introduced but can be summarized into the following categories. First, by introducing \textit{data} into the ABM, including: stylized numbers summarizing complex trends, numbers carrying specific meaning (e.g., the actual price of a given good, distance between people \cite{predhumeau2021simulating}), statistical data (e.g., census data, household distribution), and high-resolution data (e.g., high-resolution geographical data, recordings of human voices, 3D models of virtual humans). The use of multiple sources of data is sometimes argued as beneficial for realism. 
Second, by introducing \textit{insights} that may arise from common knowledge and specific theories (e.g., psychological theories). These insights can also be expressed by seeking to avoid introducing qualitative properties into the model that are considered unrealistic.
Third, through the introduction of specific \textit{mechanisms}, features, and/or model properties that are argued to enhance the realism of the ABM, including: the use of ABMs in general, specific deliberation methods (e.g., CLARION \cite{sun2004simulating}, path planning), specific Artificial Intelligence methods (e.g., reinforcement learning), specific interaction methods (e.g., social force model, group avoidance behaviors \cite{predhumeau2021simulating}), or the inclusion of certain phases of a process \cite{sinclair2023hybrid}. 
At a meta level, realism is argued to derive from the inclusion of abstract properties about models, such as agent heterogeneity, simple models, succinct models, mathematical models, complex models, models that include context-specific elements (e.g., delays between trading operations \cite{vyetrenko2020get}, use of dikes in a flooding model \cite{chapuis2019agent}), and the use of multiple models.
The vast majority of papers do not rely on a theoretical foundation to argue why the proposed method is likely to make the ABM more realistic, except for one instance (n=1) \cite{sakurahara2020enhancing}.

Summing up: the set of methods argued to making ABMs realistic appears to be highly diversified, except for a few recurring methods. The use of these methods is seemingly case-sensitive, and not grounded in theoretical foundations and overall loosely described (a few mentions of realism per paper).

\subsection{Assessing realism}

Out of the analysed documents, 40\% (n=29) explicitly demonstrate the realism of their models (however, while this number appears to be low, it may be assumed that, in general, realism is assessed implicitly through the validation step in the modelling process). Common methods mentioned for assessing the realism of a model include: \textit{proximities with quantitative data} (e.g., stylized facts, meaningful data such as specific prices, general statistical data such as demographics), \textit{compliance with high-resolution data} (e.g., comparison against video footage), \textit{compliance with qualitative expectations} (e.g., expected displayed behaviors and global dynamics, absence of unrealistic behaviors), and \textit{compliance with common-sense expectations} (e.g., unstructured visual assessment of the system).
Whereas sharing similarities with validation, the demonstration of the realism of the simulation is usually very brief (n=1). In the majority of the papers, the validation section makes no reference to realism.

\section{Discussion}

\subsection{RAINO: Reality Anchor, Input and Output}

Summing up the results, the current ABMs literature explicitly mentioning realism is loosely defining and framing realism as a common-sense concept. Yet, the meaning of realism, how it is achieved and how it is demonstrated, can vary significantly from one paper to the next (e.g., \cite{day2006challenges,sakurahara2020enhancing}).

For helping to navigate the diversity surrounding the concept of realism, this current paper introduces a framework called \textit{Reality Anchor, Input and Output}, or, in short, RAINO. The framework synthesizes the common approaches for making and for arguing that a model is realistic, building over a compilation and integration of the arguments used in the literature analyzed in this current SLR and the frames proposed by \cite{frenchrealism,giardini2023modeling}. 

The first element of the RAINO Framework is the \noindent\textbf{Reality Anchor (RA)}. RA is defined as the specific object that is claimed to reflect, depict, or capture the real world. Common instantiations of RA include: personal experience and common sense \cite{voloshin2015towards}, expert knowledge \cite{demir2012new}, theories \cite{sun2004simulating}, processes and mechanisms \cite{szilagyi2001solutions}, prior models \cite{sun2004simulating}, and data \cite{ipto2008realistic}, including high-resolution data (e.g., high-resolution Geographic Information System (GIS) data) \cite{chapuis2019agent,galea2012uk,gathmann1998inter}.

RAs can be associated with a variety of aspects, commonly influencing ones being the \textbf{strength} of the anchor, defining how easily this anchor can be discounted (e.g., amount of validation of a theory, discrepancies in the data), and the \textbf{specificity/range} of the anchor, detailing how generalized or context-specific an anchor is (e.g., specific information about a specific object in a specific situation such as "camera feeds show that door 17 was opened at 14:47" \cite{chapuis2019agent,galea2012uk} versus a generalized description of reality such as "exposure to anxiety leads to increased occurrences of avoidance and information-seeking behaviors" \cite{horned2023models}).

The second element of the RAINO Framework is \textbf{Input/Output} (INO) positioning of the RA. INO qualifies whether a given RA is used on the \textit{input} of the ABM\footnote{Input involves a broad definition, from setting the values of specific variables to the concepts, code, and underlying processes modelled by the ABM.} (e.g., integrating this anchor within design decisions and instantiations of model variables) \cite{tsai2008epidemic} or on the \textit{output} of the ABM (e.g., proximity of execution traces with the RA \cite{alfarano2007minimal}). This distinction matters because, while both input and output RAs are used for claiming the realism of various ABMs in the retrieved literature, the practical steps for relating the RA to the ABM are significantly different (for example, integrating variables from a theory in an ABM versus investigating the traces of the ABM results).

The relation between the RAs, their aspects, and their INO positioning are prime elements used for assessing the realism of a given model. A given RA can be a \textit{positive} or \textit{negative} indicator of realism \cite{khalil2022realistic}, depending on whether the model is attributed to resemble or diverge from the RA (e.g., agents eating food can be expected to be a positive indicator of realism whereas agents not aging has a negative valence over common-sense RAs). Multiple RAINO arguments may \textit{contradict} each other (e.g., data state that the population should decline whereas theories state that the population should increase) \cite{brown2008darwinian,hluchy2011handling}. The resolution of such discrepancies is usually sought by assessing the strength and specificity of the involved RAs (e.g., the data only covers one geographical area and part of the population may have migrated).

Realism assessments are also sensitive to context-specific \textit{framings of reality}, which may impact the set of considered RAs and their aspects. These framings commonly include: how assessors\footnote{In this paper, ``assessor" refers to any person engaged in appraising the realism of the model without any requirement on a formal role or competence. A modeller, a stakeholder ordering a model, a bystander in a fair, a spouse glancing over the computer can all emit an assessment of the model's realism.} \textit{generally experience} and what they consider as reality \cite{bourdieu2018cultural} (e.g., a factory worker may give more strength to RAs tied to logistics whereas a workers' union leader may give more strength to RAs about power relations; researchers from different disciplines may disagree on how strong and specific a given theory is); \textit{contextually restricted} framing of reality as identified by \cite{giardini2023modeling} (e.g., restricting the assessment to psychological realism, hence attributing less realism to the inclusion of market mechanisms); and \textit{contextually altered} framing of reality, in which the assessor considers a partly alternate reality and hence different RAs (e.g., elvish agents not aging would create realism in a model of Tolkien's fiction).

As a general use of the framework, while each RAINO argument is taken independently, ABM studies may feature multiple RAINO arguments. Likewise, a given source of data may be involved in multiple reality anchors. For example, GIS may be used both as an input RA (e.g., by laying out passable areas) and as an output RA (e.g., giving evidence of the congestion over the course of the day that the system should reproduce).

\subsection{Why RAINO?}

The concepts deployed by RAINO capture the main argumentative structures identified in the papers analyzed in the last step of the SLR: authors show the realism of their models is realistic by A) identifying a property pertaining to reality (the RA) and relating it to either B) the model input (e.g., model's structures, processes, variables values) or to C) the model's output (e.g., traces, statistical distributions, properties).

Besides capturing key discourses about realism in Agent-Based Modelling, RAINO allows elevating the discussions from current implicit dictionary definitions and assumptions (e.g., realism as an objective, mono-dimensional, terminal construct) towards considering realism as a multi-dimensional, context-sensitive construct, thus enabling the elaboration of practical methods for designing and proving models to be realistic, and providing essential concepts for framing what allows the creation and claim of realism in ABMs.

At its core, RAINO disentangles two concepts that are usually overlapped when talking about realism: \textit{the "real world" reality} (as an absolute, external object) and \textit{secondary accounts of reality} (e.g., data, theories, assumptions about what is real), which are the only objects accessible when arguing for the realism of a model and are imperfect approximations of what is considered real along certain perspectives. Hence, RAINO allows expanding discussions around realism as follows. 
First, through the possibility of qualifying various RAs, RAINO allows identifying \textit{the nature of the proximity of these RAs to reality and the conditions under which a model loses its proximity to reality}. For example, a theory-driven anchoring is commonly expected to be closer to reality when considering how individuals decide and interact, while being more susceptible to drifting away from reality over time compared to, for example, data-driven models. 
Second, by identifying and making explicit the relations between how realistic a model is assessed to be, reality, RAs, and the assessors of realism and their dispositions when assessing, RAINO allows explaining how different assessors may disagree on whether, why, and how much a model is realistic. 
Third, by providing key elements for practical questions involved in ABM development, including identifying objective factors for arguing for the realism of various RAINO arguments, as well as practical ABM design and validation questions (e.g., principled approaches for addressing common challenges of dealing with discrepancies between common sense, theories, and data). 
Fourth, by enabling the capture of dynamic properties about how people change how they think about the realism of a model, whether through short-term internal changes in perspective (e.g., intentionally focusing on certain parts of the model \cite{giardini2023modeling}) or long-term, possibly implicit revisions of one's reality (e.g., getting increasingly convinced that a model is realistic from engaging with it or with its underlying principles \cite{edmonds2017different,kuhn1997structure}). 
Fifth, by providing conceptual tools for capturing how people create and assess the realism of models despite their acknowledged fictional nature.

\subsection{RAINO explained on a practical issue}

As a practical example, it is common to show generic prototypes to stakeholders in order to elicit discussions (e.g., a generic grid-based traffic model to a city representative). Although such a prototype may elicit some interest, stakeholders commonly find these prototypes unrealistic. As to enable an effective exploration of the prospects before engaging into an in-depth development of a new model, it is common to make minor tweaks to the initial model towards fitting the stakeholder's specific case (e.g., moving the nodes and overlaying a map of the specific city onto the model's interface), even if, by doing so, clearly unrealistic features are added (e.g., agents following physics-breaking trajectories and speeds). Yet, involved stakeholders tend to see such a model as much closer to reality. This example shows a seeming paradox: depending on who you ask, the updated model is considered either more or less realistic than the original one.

The RAINO framework, by introducing an assessor-centered perspective on how realism is perceived, provides a structured framing for explaining what would be paradoxical in the intuitionist understanding of realism (i.e., realism as an exclusive attribute of the model). In this example, including the map in the model creates a "high-resolution data" input RA. Stakeholders, who are closely affiliated with the reality tied to the city/map, are more prone to associate a high strength and valence to this anchor, hence making the updated model more realistic to them than the original one. Conversely, modellers, who are more attached to conceptual soundness and accurate representations of physical and social dynamics, would consider an additional, implicit, anchor: "expectations about how fast people travel in a city" arising from common-sense, which is an output RA, and would be prone to attach an important strength to this anchor and being less sensitive to the shape of the city. Due to the updated model being further away from the original one to produce the reality depicted by this RA, the modeller sees the updated simulation as less realistic compared with the original one.

\subsection{RAINO and related approaches}

RAINO integrates and expands the insights developed in the related literature as follows. RAINO relates to \cite{frenchrealism} by expanding the range of types of RAs beyond data-driven RAs and integrating the levels of proximity of reality as part of the \textit{specificity/range} aspect. RAINO integrates and generalizes the dimensions of realism from \cite{giardini2023modeling} (e.g., psychological, behavioral, social) through its \textit{perspective} aspect. RAINO addresses the observation that data collected outside of a crisis may become irrelevant during a crisis \cite{giardini2023modeling} through its specificity/range variable.

\vspace{0.2cm}

\noindent\textbf{RAINO \& calibration, validation, verification, and empirical grounding.} Calibration, validation, and empirical grounding are key Agent-Based Modelling practices that conceptually relate to realism, albeit their crossing remains unexplored by former work. While an exhaustively crossing would extend beyond the scope of this paper, RAINO provides new perspectives for understanding their relation.

\noindent \textit{\textbf{Calibration}} is defined as "running the model with different parameters and testing, for each case, how well the model performs by comparing the output against empirical data. The goal is to find parameter sets that minimize the model’s error and can be used to provide a range of predictions or analyses" \cite{mcculloch2022calibrating}. Calibration can be invoked either in support of or against claims of realism of the model. The results of the current SLR indicate that calibration is commonly used to argue for the realism of the ABM. The proximity between system output and empirical observation is a straightforward output reality anchor. As a surprising observation that can be made through the lens of RAINO, the calibration process and outcome may be argued either positively and negatively in regards to realism depending on  the assessor and the context: on the one hand, calibration can be argued to be a systematic search for (undocumented) model parameters that are closer to reality; on the other hand, calibration may be argued  to lower realism due to the procedure focusing on optimizing towards output proximity metrics, possibly losing touch with reality (e.g., the optimal replication of market dynamics yielding to associating odd prices to common items).

\noindent \textit{\textbf{Validation}} is defined as "the process of determining if a model adequately represents the system under study for the model’s intended purpose" \cite{collins2024methods}. In this definition taken in its broadest sense, the adequateness of the representation and the model's intended purpose can disregard any relation to reality. For example, numerous models in NetLogo's library are highly valid as technical or aesthetically pleasing demonstrations, with no aim to relate to reality. Narrowing down to models that are directed towards replicating reality (e.g., prediction or explanation from the taxonomy of \cite{edmonds2017different}), this framing of validation, by focusing on adequate representations, arguably relates validation with the activity of creating input reality anchors towards allowing the assessment of realism.
The practical use of the concept of validation identified in this current SLR and in practice includes activities that may diverge from the definition of \cite{collins2024methods}, by focusing instead on establishing the proximity between the model's output and empirical data, such as comparing the output of the model against specific data points, trends of data, assumptions from theories, and feedback from stakeholders \cite{castella2005participatory}. 
In this case, validation relates to the creation of output reality anchors, may it be of a positive valence (if model output and data are argued to align) or a negative valence (if model output and data are argued to diverge).
\textit{\textbf{Verification}} is defined as "ensuring that the computer program of the computerized model and its implementation are correct" \cite{sargent2010verification}.
Verification can be argued to be a process seeking to introduce input reality anchors by deriving the realism of the implemented model from the realism associated to the original/conceptual model.
\textit{\textbf{Empirical grounding}} is defined as: "connecting model and target system, through giving values to the set of parameters in order to enable simulation" \cite{scheller2019review,smajgl2014empirical}. This definition directly relates empirical grounding to the process of creating input reality anchors.

\vspace{0.1cm}

\noindent\textbf{RAINO \& KISS and KIDS.}
\textbf{\textit{KISS (Keep it Simple, Stupid)}} is a modelling strategy that prescribes selecting the simplest possible mechanisms to generate the target phenomenon and adding complexity only if simpler models do not yield satisfactory results \cite{edmonds2004kiss}. Although no theoretical boundaries connect KISS to realism, the practice of the KISS strategy commonly involves trade-offs between model complexity and realism. In practice, KISS strategies are expected to involve the use of a minimal set of reality anchors that maximally compress a complex reality into a limited set of variables and processes. As such, the KISS approach can be expected to primarily rely on theory-based and model-based anchoring as input. In regard to output anchoring, in principle, KISS does not prescribe any specific way to demonstrate realism. Due to the simplicity of KISS models and their limited ability to fit specific cases, they can be expected to be grounded primarily in theory-based and loose data-driven anchors (e.g., general trends), both in input and output.

\textbf{\textit{KIDS (Keep it Descriptive, Stupid)}} is a modelling strategy that prescribes maximizing the acquisition and integration of as many varied sources of information within the model as possible, regardless of the resulting complexity, and only afterwards considering reducing the complexity of the model \cite{edmonds2004kiss}. The KIDS approach takes a maximalist perspective on reality anchoring, promoting the acquisition and direct integration into the model of as much anchoring as possible from a broad variety of data sources, both as input and output. However, the many sources of information may contradict each other and involve extensive efforts to resolve discrepancies in arguing for the realism of the model.

RAINO allows arguing that KISS may be a suitable approach for models to be perceived as realistic by theorists and modellers, by promoting the inclusion of anchors that relate directly to their framing of reality, particularly if they specialize in the simulated field. Conversely, the same is valid for KIDS, by including extensive anchors related to the experienced realities of stakeholders, is plausibly more realistic for these stakeholders. This crossing, which generalizes the example provided in the previous section, illustrates the relevance of considering realism along broader frames, as it can entail radically different approaches for designing realistic models and demonstrating their realism.

\vspace{0.2cm}

\noindent\textbf{Ethical implications.}
The RAINO Framework, by bringing into light the plurality of views over reality and making them explicit, provides key structures for further specifying and addressing ethical considerations related to social simulation. As examples, seeking to conform to the reality of a given stakeholder may harm conformity with other anchors (e.g., external data) and further entrench the stakeholder into an increasing disconnection with others' realities (e.g., by confirming erroneous assumptions) \cite{edmonds2017different,kuhn1997structure}. Conversely, failing to conform to a stakeholder's reality may damage collaborations and expose the stakeholder to harm, such as by challenging their worldviews (possibly leading to erroneous understandings of reality) and their social position \cite{edmonds2017different,kuhn1997structure}.

\section{Conclusion and future work}

This study engages in an in-depth analysis of 73 papers, identifying how they define, frame, and argue for attempts to create or to demonstrate the realism of ABMs. Overall, \textbf{explicit reflection over realism as a concept appears to be in its infancy}: realism is broadly undefined, unframed, and used presumably based on a vague dictionary definition. Dozens of radically different methods have been developed for creating and validating realism and, despite this variety, there is no structured approach to argue that a model is indeed realistic. While, arguably, realism may be used performatively to express a proximity between the model and assumed reality rather than grounded in dedicated conceptualizations and assessment methods, realism remains nonetheless a ubiquitously used concept in ABM research and practice. Incidentally, the development of a structured conceptualization of realism, may it be to identify its limitations as a concept and thus refine more suited concepts or for developing new methods for producing and demonstrating model realism, is essential to support the effective continued development of ABM research and practice.

To address this gap, this paper introduced the Reality Anchor, Input, Output (RAINO) Framework, which systematizes the arguments of how realism is created and integrated in ABMs. RAINO expands the current framing of realism, in which ABMs can be more or less realistic along some absolute scales, by accounting for the relational nature of realism, where reality anchors (e.g., data, theories) are used to show the proximity of the model with reality and where the assessment of these anchors may be subject to divergences from one assessor and assessment context to the next. RAINO is then demonstrated to be suitable for explaining how various classic modelling strategies (e.g., KISS and KIDS) may elicit various assessments of realism by various categories of stakeholders.

The application of the SLR method brings  several limitations, including possibly missing some references due to the restrictions set by our query and by the database \cite{page2021prisma} (e.g., some papers may discuss in length about realism despite not mentioning realism in their title). The relatively limited discussions of realism within the retrieved papers, while already a significant result, also limits the amount of material available for building up and designing the RAINO Framework. 
The set used as input for the in-depth qualitative analysis step of the SLR can be expanded with additional sources of information, such as all papers featuring realism not only in their titles but also in their abstract and keywords (n=2193 documents) and documentation schemes attached to models (e.g., ODD protocol \cite{grimm2020odd}, RAT reporting scheme \cite{achter2022rat}). Due to the scaling up of prospective data, different analysis methods, possibly inherited from natural language processing, would be required for making such an analysis feasible.

The RAINO Framework and its novel conceptual foundations for reasoning about realism in an Agent-based Modelling context open as future work: expanding RAINO with additional concepts; 
exhaustively listing how RAINO variables/arguments are operationalized (e.g., an atlas of all methods used as Input Anchors);or developing design and assessment methods for supporting the integration and demonstration of realism in ABMs, together with the application of these methods on practical (real-world) use-cases.

\section*{Acknowledgments}

The authors thank the reviewers for their constructive suggestions. 
This research has been conducted within the frame of the Anxiety-Sensitive AI initiative (AnxSAI)\footnote{\url{https://www.umu.se/en/research/projects/anxsai-anxiety-sensitive-artificial-intelligence/}}. The first author thanks Bruce Edmonds and Maarten Jensen for early discussions related to the contents of this paper. The second author thanks Sebastian Achter for earlier discussions about world representation in Agent-Based Modelling.
The first author acknowledges the financial support of the project ``Anxiety-Sensitive Artificial Intelligence'', funded by the Swedish Research Council (project number 2023-04505), Sweden. 
The second author acknowledges the financial support of the project ``FUTURES4Fish: Adaptive socio-technological solutions for Norwegian fisheries and aquaculture'', funded by the Research Council of Norway (project number 325814).

 \bibliographystyle{splncs04}
 \bibliography{biblio}
\newpage

\section*{Annexes}

 \begin{longtable}{p{12cm}|l}
    \caption{List of all papers retained for in-depth analysis step of the Systematic Literature Review performed for this study}        
    \label{table:all_papers}
  \\
    \textbf{Title} & \textbf{Reference}  \\\hline
         Inter-site: A new tool for the simulation of spatially realistic population dynamics & \cite{gathmann1998inter}\\
Evolution of migration rate in a spatially realistic metapopulation model & \cite{heino2001evolution}\\
Solutions to realistic prisoners' dilemma games& \cite{szilagyi2001solutions}\\
Towards biologically realistic estimates of home range and spatial exposure for colonial animals&\cite{niven2025towards}\\
On building an organizationally realistic agent-based model of local interaction and emergent network structure& \cite{hazy2004building}\\
Challenges of biological realism and validation in simulation-based medical education&\cite{day2006challenges}\\
Beyond cellular automata, towards more realistic traffic simulators&\cite{correia2006beyond}\\
How realistic should knowledge diffusion models be?&\cite{cointet2007realistic}\\
Simulating organizational decision-making using a cognitively realistic agent model&\cite{sun2004simulating}\\
Realistic spatial backcloth is not that important in agent based simulation: An illustration from simulating perceptual deterrence & \cite{elffers2008realistic}\\
Darwinian fisheries science needs to consider realistic fishing pressures over evolutionary time scales&\cite{brown2008darwinian}\\
Epidemic dynamics and thresholds in agent-based simulations under realistic resources and cost conditions&\cite{tsai2008epidemic}\\
Realistic agent populations for large-scale virtual training environments&\cite{ipto2008realistic}\\
EpiSimdemics: An efficient algorithm for simulating the spread of infectious disease over large realistic social networks&\cite{barrett2008episimdemics}\\
The impact of realistic age structure in simple models of tuberculosis transmission&\cite{brooks2010impact}\\
An iterative approach for generating statistically realistic populations of households&\cite{gargiulo2010iterative}\\
How much time can herbivore protection buy for coral reefs under realistic regimes of hurricanes and coral bleaching?&\cite{edwards2011much}\\
Handling internal complexity in highly realistic agent-based models of human behaviour & \cite{hluchy2011handling}\\
Agent's minimal intelligence calibration for realistic market dynamics&\cite{veryzhenko2010agent}\\
Key points for realistic agent-based financial market simulations&\cite{veryzhenko2011key}\\
A new driver behavior model to create realistic urban traffic environment&\cite{demir2012new}\\
Implementing comprehensive offender behaviour in a realistic agent-based model of burglary & \cite{malleson2012implementing}\\
Simulating the spread of infectious disease over large realistic social networks using Charm++&\cite{bisset2012simulating}\\
A minimal noise trader model with realistic time series properties&\cite{alfarano2007minimal}\\
Agent-based large scale simulation of pedestrians with adaptive realistic navigation vector fields& \cite{karmakharm2010agent}\\
How level of realism influences anxiety in virtual reality environments for a job interview&\cite{kwon2013level}\\
From real purchase to realistic populations of simulated customers&\cite{mathieu2013real}\\
Incorporating toxicokinetics into an individual-based model for more realistic pesticide exposure estimates: A case study of the wood mouse&\cite{liu2014incorporating}\\
Towards realistic and effective Agent-based models of crowd dynamics&\cite{wkas2014towards}\\
Modelling hen harrier dynamics to inform human-wildlife conflict resolution: A spatially-realistic, individual-based approach&\cite{heinonen2014modelling}\\
Development of an Evaluation Method for Artificial Market Settings Considering a Realistic Pricing Mechanism&\cite{mizuta2012development}\\
Towards a performance-realism compromise in the development of the pedestrian navigation model&\cite{voloshin2015towards}\\
Structural realism, emergence, and predictions in next-generation ecological modelling: Synthesis from a special issue&\cite{grimm2016structural}\\
The importance of realistic dispersal models in conservation planning: application of a novel modelling platform to evaluate management scenarios in an Afrotropical biodiversity hotspot&\cite{aben2016importance}\\
Cellular Automata as the basis of effective and realistic agent-based models of crowd behavior&\cite{lubas2016cellular}\\
Recalibrating disease parameters for increasing realism in modeling epidemics in closed settings&\cite{bioglio2016recalibrating}\\
Realistic human behaviour simulation for quantitative ambient intelligence studies&\cite{veronese2016realistic}\\
Driving behaviour clustering for realistic traffic micro-simulators&\cite{petraro2017driving}\\
Enhancing the realism of simulation (EROS): On implementing and developing psychological theory in social simulation&\cite{jager2017enhancing}\\
Massively parallel simulations of spread of infectious diseases over realistic social networks&\cite{bhatele2017massively}\\
Using realistic trading strategies in an agent-based stock market model&\cite{llacay2018using}\\
Generating Realistic road usage information and origin-destination data for traffic simulations: Augmenting agent-based models with network techniques&\cite{hofer2017generating}\\
Feed-in tariffs for solar microgeneration: Policy evaluation and capacity projections using a realistic agent-based model&\cite{pearce2018feed}\\
The importance of modelling realistic human behaviour when planning evacuation schedules&\cite{bulumulla2017importance}\\
Generation of realistic mega-city populations and social networks for agent-based modeling&\cite{burger2017generation}\\
Realistic agents with social practices&\cite{mercuur2018realistic}\\
The ODD protocol for describing agent-based and other simulation models: A second update to improve clarity, replication, and structural realism &\cite{grimm2020odd}\\
Norms in Social Simulation: Balancing Between Realism and Scalability&\cite{pastrav2020norms}\\
Integration of Genetic Algorithm and Agent Based Model to Visualize Near Realistic Sustainable Urban Growth: A Comparative Study&\cite{chandan2020integration}\\
Gen*: An Integrated Tool for Realistic Agent Population Synthesis&\cite{chapuis2019gen}\\
Towards a more realistic simulation of public transit: Generating transit schedules with vehicle circulations&\cite{marburger2021towards}\\
Enhancing realism in fire probabilistic risk assessment of nuclear power plants&\cite{sakurahara2020enhancing}\\
Get real: Realism metrics for robust limit order book market simulations&\cite{vyetrenko2020get}\\
An Agent-Based Co-modeling Approach to Simulate the Evacuation of a Population in the Context of a Realistic Flooding Event: A Case Study in Hanoi (Vietnam)&\cite{chapuis2019agent}\\
A hybrid data gathering and agent based cognitive architecture for realistic crowd simulations&\cite{sinclair2023hybrid}\\
Simulating realistic pedestrian behaviors in the context of autonomous vehicles in shared spaces&\cite{predhumeau2021simulating}\\
Realistic agent-based simulation of infection dynamics and percolation&\cite{nagel2021realistic}\\
Intellectual Route Planning Methods for Realistic Agents' Movement&\cite{chebotkov2019intellectual}\\
High Performance Agent-Based Modeling to Study Realistic Contact Tracing Protocols&\cite{hoops2021high}\\
An economically realistic asset exchange model&\cite{boghosian2022economically}\\
Towards More Realism in Pedestrian Behaviour Models: First Steps and Considerations in Formalising Social Identity&\cite{wijermans2022towards}\\
Increasing the realism of electricity market modeling through market interrelations&\cite{qussous2022increasing}\\
Confronting spatial capture–recapture models with realistic animal movement simulations&\cite{theng2022confronting}\\
Realistic urban traffic simulation with ride-hailing services: a revisit to network kernel density estimation (systems paper)&\cite{khalil2022realistic}\\
Fit for purpose: Modeling wholesale electricity markets realistically with multi-agent deep reinforcement learning&\cite{harder2023fit}\\
Agent-based model for realistic vessel route simulation in port areas&\cite{gueli2023agent}\\
Comparing different approaches of agent-based occupancy modelling for predicting realistic electricity consumption in office buildings&\cite{mashuk2024comparing}\\
Multi-Agent Pathfinding with Obstacle Movement for Realistic Virtual Tactical Simulations on Topographic Terrains&\cite{souza2023multi}\\
Modeling Realistic Human Behavior in Disasters. A Rapid Literature Review of Agent-Based Models Reviews&\cite{giardini2023modeling}\\
Reinforcement Learning in Agent-Based Market Simulation: Unveiling Realistic Stylized Facts and Behavior&\cite{yao2024reinforcement}\\
Assessing the behavioral realism of energy system models in light of the consumer adoption literature. & \cite{ball2025assessing}\\

\end{longtable}

\end{document}